\input epsf
\documentclass[nohyper,notoc]{article} 

\usepackage{amsmath}
\usepackage{amssymb}
\usepackage{axodraw}
\usepackage{epsfig}
\usepackage{bm}

\def\beq{\begin{equation}}
\def\eeq{\end{equation}}
\def\bea{\begin{eqnarray}}
\def\eea{\end{eqnarray}}
\def\bq{\begin{quote}}
\def\eq{\end{quote}}
\def \lsim{\mathrel{\vcenter
     {\hbox{$<$}\nointerlineskip\hbox{$\sim$}}}}
\def \gsim{\mathrel{\vcenter
     {\hbox{$>$}\nointerlineskip\hbox{$\sim$}}}}
\def\gappeq{\mathrel{\rlap {\raise.5ex\hbox{$>$}}
{\lower.5ex\hbox{$\sim$}}}}
\def\lappeq{\mathrel{\rlap{\raise.5ex\hbox{$<$}}
{\lower.5ex\hbox{$\sim$}}}}

\def\ETm{E_T \!  \! \! \! \! \! \! /~~}

\def\Dlr{\overset{\leftrightarrow}{D^\nu}}

\def\m3e{\mu \to e \bar{e} e}
\def\a{\alpha}

\def\g{\gamma}

\def\m{\mu}

\begin{document}

\renewcommand{\thefootnote}{\fnsymbol{footnote}}
\begin{center}
{\Large {\bf 
Including the Z in an Effective Field Theory for dark matter
at the LHC}}
\vskip 25pt
{\bf   Sacha Davidson $^{1,}$\footnote{E-mail address:
s.davidson@ipnl.in2p3.fr}   }

\vskip 10pt  
$^1${\it IPNL, CNRS/IN2P3,  4 rue E. Fermi, 69622 Villeurbanne cedex, France; 
Universit\'e Lyon 1, Villeurbanne;
 Universit\'e de Lyon, F-69622, Lyon, France
}\\
\vskip 20pt
{\bf Abstract}
\end{center}
\begin{quotation}
  {\noindent\small 
An Effective Field Theory for dark matter at a
TeV-scale hadron collider  should include contact interactions
of dark matter with the partons, the Higgs and
the $Z$. This note estimates the impact of 
including dark matter-$Z$ interactions on the complementarity
of spin dependent direct detection and
LHC monojet searches for dark matter. The
effect of the $Z$ is small, because  it interacts
with  quarks via small
electroweak couplings, and 
the contact interaction self-consistency condition
$C/\Lambda^2 < 4\pi/\hat{s}$ restricts the coupling to dark matter.
In this note, the contact interactions between the $Z$ and
dark matter are parametrised by derivative operators;
this is convenient at colliders because such interactions
do not match onto low energy  quark-dark matter contact
interactions.

\vskip 10pt
\noindent
}

\end{quotation}

\vskip 20pt  

\setcounter{footnote}{0}
\renewcommand{\thefootnote}{\arabic{footnote}}

\section{Introduction}

Various experiments attempt to detect the particle
making  up the  ``dark matter''\cite{DM} of our Universe. 
 For instance,
direct detection(DD) experiments \cite{EdelCDMS,Xenon,SC,autres}, 
search  for $\sim $ MeV energy deposits
due to scattering of dark matter particles from
the galactic halo on detector nuclei.  And 
the Large Hadron Collider (LHC)
searches\cite{CMS,ATLAS} for dark matter pairs
produced in multi-TeV $pp$ collisions,   which
would materialise  as an excess of events
with  missing energy and  jets.
The LHC and DD searches are at very different energy
scales, so different Standard Model (SM) particles are present,
and also the quantum interferences are different\cite{PST}. 
The expected rates can be compared in specific 
dark matter models \cite{R}, or, in recent years,
several studies\cite{CMS,tevatron,toutlmonde,Z,unitarity1,monoH}  
have compared the
LHC and DD sensitivities using  
a contact interaction parametrisation  of 
the dark matter  interactions with the
standard model particles.

The LHC  bounds obtained   in this way are restrictive,
and probe smaller couplings than  
 direct detection
experiments searching for ``spin dependent''
interactions between partons and dark matter
\cite{SC}.
 These contact
interaction studies are refered to as 
``Effective Field Theory'' (EFT), and 
 considered to be relatively model independent.
However, the particle content
is an input in EFT, and the restrictive LHC  limits  
assume that the dark matter
particle is the only new particle
accessible at the LHC. Relaxing this assumption
can significantly modify the experimental
sensitivities\cite{st,pvz,toni2}.
This has motivated various simplified models
for dark matter searches at the LHC \cite{marcusmodel,DMLQ,deSGS}.
Retaining this assumption, as will
be done in this note, is only
marginally consistent, because the contact interactions 
to which the LHC is sensitive
would have to be mediated by strongly coupled particles.
As recalled in the next section, this implies  
that colliders can exclude contact interactions
of order their sensitivity, but not  much larger.

Effective Field Theory (EFT) is  a recipe to get the correct answer in a 
simple way\cite{Georgi}.  So this note attempts 
to compare LHC and DD constraints on dark matter,
according to the prescriptions of
\cite{Georgi}.
An EFT for dark matter at the LHC
 should parametrise all  possible
 SM-gauge invariant interactions of the dark matter with 
other on-shell particles.
So first,
 contact interactions between the dark matter  
 and the Higgs or $Z$ should be included  at the LHC.
 These can  interfere with the contact interactions
studied in previous analyses, 
 but contribute differently at colliders
from  in direct detection,
so the linear
combination of operator coefficients constrained
at high and low energy will be  different.
Secondly, an EFT contains in principle a tower
of operators\cite{Porod} organised in  increasing powers of the
inverse cutoff scale $1/\Lambda$, 
and higher orders  can  only be neglected if there
is a sufficient hierarchy of scales: $\Lambda_{NP}\gg v$.
This hierarchy is absent in dark matter production
at the LHC. Addressing  
the  importance of higher dimensional operators will be
left to a subsequent publication
\footnote{ Higher dimensional operators can contain more
fields and be  suppressed
by phase space, or contain Higgs fields and
be suppressed by $\langle H \rangle^2/\Lambda^2$,
or contain derivatives and be dangerous. }. 

This note focuses on including  the $Z$ in
the EFT for dark matter at the LHC,
and  estimates analytically 
the consequences of including the lowest dimension
operators allowing dark matter interactions with the
 $Z$ \footnote{Contact interactions between dark matter and the  $Z$
have been  proposed  in \cite{deSGS}   as a 
benchmark model, assuming other contact interactions
to be absent. }. Section \ref{sec:EFT} outlines
a  peculiar  choice of 
operators  for the $Z$ vertex; 
they are proportional to the momentum-transfer-squared.
This choice appears convenient, because the effects of the $Z$
are  therefore absent in direct detection.  
Section \ref{sec:LHC}  estimates the impact
of  cancellations between $Z$ exchange and 
dark matter contact interactions with quarks
at the LHC, and section \ref{sec:DD} recalls the direct detection
bounds.

\section{EFT, assumptions and  operators  }
\label{sec:EFT}

The low energy consequences of New Physics from above
a scale $\Lambda$ can be parametrised by contact interactions
of coefficient
$C/\Lambda^n$. Unitarity \cite{unitarity1,unitarity2}
approximately implies that  $C< 4\pi$,
 and the contact interaction approximation
implies that the  momentum exchange  should be less than  $\Lambda$.
This means that  an experiment can {\it exclude}
\beq
\frac{4\pi}{ \hat{s} } > \frac{C}{\Lambda^2} > {\rm sensitivity}~~,
\label{plage}
\eeq
where $\hat{s}$ is the four-momentum-squared of the process.
Low energy experiments, where $\hat{s} \to 0$, therefore
can be taken to exclude everything above their
sensitivity.  However, the upper limit of  
eqn (\ref{plage}) is relevant 
for  collider searches, where $\hat{s}$
is the invariant mass of the invisibles.
This upper limit is rarely taken into account in the literature.

The  first step in the EFT recipe to parametrise New Physics
from  beyond the scale $\Lambda$
it to add to the
Lagrangian (at the scale  $\Lambda$),
 all the  non-renormalisable  operators
 which can be constructed
out of the fields present, consistently with the symmetries
of the theory\cite{Georgi}. The coefficients $C^{(n)}_O$ of
these operators are unknown ``coupling constants''
which evolve  with scale via Renormalisation Group Equations.
This infinite set of operators would be unmanageable, 
so EFT is useful when there is a hierachy 
between the  experimental and NP scales.
Then only the lowest
dimension operators need be considered.

In this note, the dark matter is  assumed to be
the only new New Physics
 particle  lighter than a TeV,  and is taken to be a SM  
gauge singlet dirac  fermion  $\chi$ with a conserved parity,
and of mass $m_\chi \geq m_Z/2$ (maybe $\geq m_h/2$), to avoid bounds on the
coupling to the $Z$ from the invisible width\footnote{For 
$m_\chi <m_Z/2$, the invisible width of the $Z$ 
(at ``2$\sigma$'', so\cite{PDB}  $\Gamma (Z\to \chi
\overline{\chi})\leq  3$ MeV) imposes that
$ | C_{Z,B}| < 8.9 (\Lambda/{\rm TeV})^2$, for 
$B= V,A$. 
} of the $Z$ (and Higgs).
So the particle content of
the EFT for $\chi$ at the LHC 
should  be $\chi$, plus all relevant particles of
the SM, which I take to be the partons, the Higgs,
and the $Z$.

The operators  should be   
SM  gauge invariant, to profit from our knowledge of the SM gauge sector.
They are  of dimension $>$ 4, and should
attach a $\chi \overline{\chi}$ pair to
partons, to the Higgs, or to  the $Z$.
The quark operators are taken  generation diagonal;
flavour-changing operators were considered in \cite{CK11}. 
The quarks are chiral because the
operators are SM gauge invariant, 
 and also
because opposite chiralities do not interfere at the LHC.
The dark matter currents are taken in a vector, axial vector,
etc basis because these do not interfere in direct
detection, nor at the LHC in the limit where the
$\chi$ mass is neglected, as done here. 

I focus on operators of lowest dimension, that is  six
 and seven. This is an arbitrary simplification, because
$\Lambda \sim$ TeV, which is the energy scale probed
at the LHC. The contact interactions considered here therefore
do not provide a ``model-independent'' parametrisation of the
interactions of $\chi$ with the SM. This problem is
left for a later publication. Concretely,
 $\Lambda$ will be taken as 1- $2 $ TeV,
for reasons discussed above eqn (\ref{lim3}).
Experimental limits on contact interactions
will therefore be presented as limits on the
dimensionless coefficient $C^{(n)}_O$.

  At dimension six,
there are  vector and 
axial vector    $\chi$ currents coupled to quarks:
\bea
\frac{C_{QX,V}}{\Lambda^2}~
\overline{\chi} \g_{\mu}\chi 
\overline{Q_i} \g^{\mu}P_X Q_i
~~~,~~~
-\frac{C_{QX, A}}{\Lambda^2}~
 \overline{\chi} \g_{\mu}\g_5\chi \overline{Q_i} \g^{\mu}P_X Q_i 
\label{qDMd6}
\eea
where the quarks  $Q_i$ are
first generation SM multiplets $   \{q_L,u_R,d_R\}$, 
and  $P_X$ is the appropriate
chiral projector.

The contact interactions between the dark matter
and the $Z$ boson are taken as
\bea
-\frac{C_{Z,V}}{\Lambda^2}
 D^\mu B_{\mu\nu}\overline{\chi} \gamma_\mu\chi
~~~\to~~~ s_{\rm w} p_Z^2 \frac{C_{Z,V}}{\Lambda^2}
 Z^\mu  \overline{\chi} \gamma_\mu\chi 
\nonumber \\
\frac{C_{Z,A}}{\Lambda^2} D^\mu B_{\mu\nu}
\overline{\chi} \gamma_\mu\gamma_5\chi
~~~\to~~~ -s_{\rm w} p_Z^2 \frac{C_{Z,A}}{\Lambda^2} Z^\mu  \overline{\chi} \gamma_\mu\g_5\chi 
\label{inteff}
\eea
where  to the right of the
arrow is the resulting  vertex, $B^\mu$ is the hypercharge gauge boson
with coupling $g'=e\tan\theta_W \equiv e s_{\rm w}/c_{\rm w}$, 
 $B^{\mu\nu} =  \partial ^\mu B^\nu -  \partial ^\nu B^\mu$,
and  a term $\propto p_Z\cdot Z$ was dropped after the arrow  in 
the axial current operator, assuming the $Z$ was produced by light quarks.
There is  in addition a ``dipole moment'' operator
$B^{\mu \nu}\overline{\chi} \sigma_{\mu \nu}\chi $, which is
neglected here because it also induces 
dark matter interactions with the photon \cite{magmoDM}
which are more interesting.

Then at dimension seven, there are four-fermion operators:
\bea
\frac{C^{(7)}_{d,S} }{\Lambda^3} \overline{\chi} \chi
~\frac{1}{2} {\Big (}  \overline{q_L} H d 
+ [\overline{q_L} H d ]^\dagger  {\Big )} 
~~~,~~~
\frac{C^{(7)}_{d,P}}{\Lambda^3}  \overline{\chi}\g_5 \chi
~\frac{1}{2} {\Big (}  \overline{q_L} H d
+ [\overline{q_L} H d ]^\dagger  {\Big )} 
\nonumber\\
\frac{C^{(7)}_{d,T} }{\Lambda^3}
\overline{\chi}\sigma^{\mu\nu} \chi 
~\frac{1}{2} {\Big (}\overline{q_L} H \sigma_{\mu\nu} d 
+ [\overline{q_L} H \sigma_{\mu\nu} d  ]^\dagger {\Big )}
~~~~~~~~~~~~~~~\nonumber
\eea
(and similarly for $u$ quarks, but with a charge conjugate
Higgs field),   interactions with the gluons:
$$
\frac{C^{(7)}_{gg,S}  }{\Lambda^3}
\overline{\chi} \chi G_{\mu \nu}^A  G^{\mu \nu, A}
~~~,~~~
\frac{C^{(7)}_{g\tilde{g},P}  }{\Lambda^3}
\overline{\chi}\g_5 \chi G_{\mu \nu}^A  \tilde{G}^{\mu \nu, A}~~~,
$$
and   double-derivative
interactions  between dark matter
and   the Higgs:
\bea
 H^\dagger D^\mu D_{\mu} H
\overline{\chi} \chi
~~~\to~~~- m_W^2 W_\mu^+W^{-\mu} \overline{\chi} \chi
-  m_Z^2 Z_\mu Z^{ \mu} \overline{\chi} \chi
+ \frac{v p_h^2 }{\sqrt{2}} h\overline{\chi} \chi
\nonumber
\\
 H^\dagger D^\mu D_{\mu} H
\overline{\chi} \gamma_5\chi 
~~~\to~~~
- m_W^2 W_\mu^+W^{- \mu} \overline{\chi} \gamma_5 \chi
-  m_Z^2 Z_\mu Z^{ \mu} \overline{\chi} \gamma_5 \chi
+ \frac{v p_h^2 }{\sqrt{2}} h\overline{\chi} \gamma_5\chi
\label{higgs}
\eea
 where $\langle H \rangle = v = 174$ GeV, $p_h$
is the four-momentum of the physical
Higgs particle $h$,  and after the
arrow are the interactions induced by the operator.

The $Z$ and Higgs  operators are choson $\propto p^2$ so that they
are relevant at the LHC  where the $Z$ and Higgs are  external
legs in the EFT,  but 
do not contribute in  the low-energy scattering
of DD.  This choice should be acceptable, because
 the operator basis can always 
be reduced by using the equations of motion\cite{Simma}.  
Focussing for simplicity on the hypercharge boson $B$,
and  neglecting gauge-fixing terms, the
equations of motion are \cite{polonais}
\beq
 D_\mu B^{\mu\nu}= g'y_H (H^\dagger D^\nu  H - (D^\nu H)^\dagger  H) +
g'\sum_\psi y_\psi \overline{\psi}\g^\nu \psi
\label{EoMZ}
\eeq
where $\psi$ is a SM fermion of hypercharge  $y_\psi$.  
 Usually\cite{polonais}, operators
containing   the double derivative on the left
of eqn(\ref{EoMZ})  are dropped, and
the operators  containing
 the Higgs v.e.v. squared $ \langle H^\dagger \Dlr  H \rangle$
are retained.  In this usual basis,
$\chi-Z$ interactions could be parametrised
by $(\overline{\chi}\gamma^\mu \chi) H^\dagger \Dlr  H$,
in which case the matrix element for      $Z$
exchange  at the LHC  is $\propto m_Z^2/(p_Z^2 -m_Z^2)$,
so negligeable for $p_Z^2 \gg m_Z^2$. 
But $Z$ exchange  should   be included in the quark-$\chi$
contact interaction used in direct detection,  
 so the coefficient of  the operators of
eqn (\ref{qDMd6}) would  not be  the same in direct
detection as at the LHC. To avoid this discrepancy,
I retain  the derivative operators of eqn 
(\ref{inteff}), and use   eq. (\ref{EoMZ})
to remove the operator 
$ \propto \langle H^\dagger \Dlr  H  \rangle$. This
means that the $Z$ couples significantly to
$\chi$ at the LHC, but negligeably  in DD, and the
operator coefficients do not change when the $Z$
is matched out of the theory.

In the case of  the Higgs, the equation of motion
is
$$
D_\mu D^\mu H = \mu^2 H -\lambda H^\dagger H H -
\overline{e}Y_e^\dagger P_L
\ell  -  \overline{d}Y_d^\dagger q_L +
\varepsilon \overline{q_L}Y_u u 
$$
where $Y_f$ are  Yukawa matrices. This has been
 used to exchange
 the more usual $ (H^\dagger H)^2 \overline{\chi} \chi$,
and $ (H^\dagger H) \overline{\chi} \chi$
operators for the  double-derivative
interactions  between dark matter
and   the Higgs given in  eqn (\ref{higgs}).
Notice that it is possible  to  use the equations of motion to
replace two operators ($\overline{\chi} \chi H^\dagger H$ and
 $\overline{\chi} \chi (H^\dagger H)^2$)
with one (involving the DM and
  $H^\dagger D^2 H$), because I am only
interested in the $h$-$\chi$-$\bar{\chi}$  interaction
induced by these operators. The  linear combination
 of operators $[\mu^2  H^\dagger H -\lambda (H^\dagger H)^2]\overline{\chi}\chi$,
which is orthogonal to the combination in the Equations of Motion,
gives a vanishing $h$-$\chi$-$\bar{\chi}$  interaction,
due to the minimisation condition of the Higgs potential.
As in the case of the $Z$, the derivative
operators of  eqn (\ref{higgs})
 are interesting, because they give 
a higgs coupling to dark matter  $\propto p_h^2$,
which has the desirable feature
of being relevant at the LHC where the Higgs is in
the effective theory, but not contributing at low energy.

This note focusses on the interactions of $\chi$
with the $Z$ (eqn  \ref{inteff}), and 
with the quark currents  of eqn (\ref{qDMd6})
which can interfere with $Z$ exchange.  So
the  dimension seven operators will be neglected in
the following sections. However, it is interesting
to first review the sensitivity to the
coefficients of the  operators of  eqn (\ref{higgs}).
The  dark matter interactions to
 $W$ and $Z$ pairs 
 were studied in \cite{HR}, who used $U(1)_{em} \times
SU(3)$ invariant operators such that
these contact interactions have  dimension five
with coupling $1/\Lambda_{CHLR}$. They find that the  8 TeV
LHC with luminosity 25 fb$^{-1}$ could probe
$\Lambda_{CHLR} \lsim $ TeV. This constrains the
coefficients of  the operators   of eqn (\ref{higgs}) 
to be $ \lsim 1/({\rm TeV}m_W^2)$, which is not
restrictive.  For $m_\chi< m_h/2$, 
a more significant limit of 
10 TeV$^{-3}$ arises  
from requiring $\Gamma(h\to \chi \overline{\chi})
\lsim \Gamma(h\to b \overline{b})$.
This restriction  should be reasonable\cite{BMM}
because the Higgs is observed to decay to $b\bar{b}$.

\begin{figure}[ht]
\begin{center}
\epsfig{file=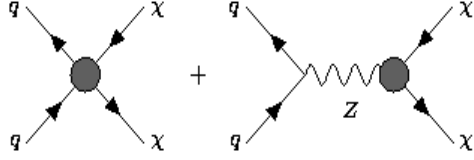, height=3cm,width=7cm}
\end{center}
\caption{Effective interactions contributing to $q\bar{q} \to
\chi \bar{\chi}$ at the LHC. The coefficient of the four fermion
operator is $C_{q,AX}/\Lambda^2 $,  and the effective axial
vector  coupling 
of the $Z$ to dark
matter is  $s_{\rm w} p_Z^2 C_{Z,A}/\Lambda^2$.}
\label{figop}
\end{figure}

\section{Estimated limits from the LHC}
\label{sec:LHC}

  Dark matter particles are invisible to
the LHC detectors,  so pair production of $\chi$s
can be searched for in  events with missing transverse energy
($\ETm)$,
which can be identified by jet(s) radiated from the
incident partons. The principle Standard
Model background for  such  ``monojet''
 searches is
$Z +$ jet production, followed by $Z\to \bar{\nu}\nu$. 
 The 8 TeV  LHC is sensitive to dark matter
contact interactions with $C/\Lambda^2 \sim $ TeV$^{-2}$.

Given the operators of eqns (\ref{qDMd6}) and 
(\ref{inteff})  at the LHC, 
the axial vector  dark matter current can interact
with  quarks $Q$ via the diagrams of figure \ref{figop}, which
 can be written as a four-fermion interaction of coefficient
\bea
c_{QX,A} & = &  C_{QX,A}+   g_X^Q \frac{gs_{\rm w}}{2c_{\rm w}}
\frac{p_Z^2 \,C_{Z,A}}{p_Z^2 -m_Z^2} 
\nonumber\\
&\underset{p_Z^2 \gg m_Z^2}{\to} &C_{QX,A}+   g_X^Q \frac{gs_{\rm w}}{2c_{\rm w}}
 C_{Z,A} ~~~,
\label{annuler}
\eea
where $g_X^Q  =\{ 1- \frac{4}{3}s_{\rm w}^2,- \frac{4}{3}s_{\rm w}^2,
 -1+ \frac{2}{3}s_{\rm w}^2, \frac{2}{3}s_{\rm w}^2\}$
for $\{u_L,u_R,d_L,d_R\}$  \cite{PDB}.
A similiar expression can be obtained  for 
the vector $\chi$  current. 
The   $Z$ exchange looks like  a  contact
interactions for large
$p_Z^2 = M^2_{inv} \gg m_Z^2$, where
$M^2_{inv}$ is the invariant mass-squared of the dark matter pair. 
This  is a useful approximation, because the 
arguments below suggests that 
most $\chi\bar{\chi}$ events 
arise at larger $M^2_{inv}$.

The aim here is to  analytically estimate
the invisible four-momentum-squared $ M^2_{inv} $,  by
comparing the partonic cross-sections for
$\nu\bar{\nu}$ and $\chi\bar{\chi}$ production. 
I assume that the QCD part of the amplitude
is identical in both cases, so it does not
need to be calculated. This allows
 for an arbitrary number of jets,
which is more difficult to simulate\cite{Uli2}
(the data frequently contains more than one jet\cite{CMS}).
In the matrix element  for jets +$\nu \bar{\nu}$
will appear 
$$
g_X^Q\frac{g^2}{4c^2_W}
\frac{1}{p^2 -m_Z^2 +im_Z\Gamma_Z}
 (\bar{Q} \g^\a P_X Q )
( \overline{\nu}\g_\a P_L\nu)
$$
whereas,  for DM production via
the  $\bar{\chi} \g^\mu\g_5 \chi$  current, this
is replaced by:
$$
\frac{c_{QX,A}}{\Lambda^2}
(\bar{Q} \g^\a P_X Q )
 (\overline{\chi}\g_\a\g_5 \chi)~~.
$$
Then the full matrix element must be squared,
and integrated over   the phase space of
 $N$ jets and  two  invisible  particles. 
The invisibles can be treated as a single
particle of variable mass $p^2 = M^2_{inv}$, using the  identity
\bea
d\Phi_{N+2} &=& \delta^4(P_{in}- \sum q_i -p) \prod_{i:1..N}
\left( \frac{d^3q_i}{2E_i(2\pi)^3} \right)
\times  
(2\pi)^3 dp^2 \delta^4(p-p_\chi - p_{\bar{\chi}})
\frac{d^3p_\chi}{2E_\chi(2\pi)^3}
\frac{d^3p_{\bar{\chi}}}{2E_{\bar{\chi}}(2\pi)^3}~~.
\nonumber
\eea
Neglecting spin correlations and the dark matter mass,
 the invisible phase space integral over
the gamma-matrix trace for  the invisible fermions gives 
$M^2_{inv}/(8\pi)$ for $\chi$s, and $3M^2_{inv}/(16\pi)$ for neutrinos.
For neutrinos in the final state,
$M_{inv}^2 = m_Z^2$
due to  the delta-function-like behaviour of the
$Z$ propagator-squared.
However, for dark matter, the $d M^2_{inv}$
phase space integral will privilege
larger values of $M^2_{inv}$.
Treating the $N$ jets of the event
as a particle of negligeable mass, the upper bound
on $M_{inv}^2$ is $ \gsim 4 \ETm^2$,
where $\ETm$ is the
invisible transverse energy. The 
CMS study \cite{CMS} uses
the range 400 GeV $\leq \ETm\lsim$  TeV. 
However,   the assumption
that  the jet emission part
of the cross-section is the same as for
$\nu$  pairs will fail, if
$M_{inv}^2$ is a significant fraction of the energy of the event.
With  the $M_{inv}$ cutoff ranging from 800 GeV  to 2 TeV, 
requiring that the
dark matter contribute $\lsim 1/6$ \cite{CMS} of
the SM background, 
gives an estimated bound   $\Lambda \gsim 880 \to 2200$ GeV,
for $c_{uL,A} =c_{uR,A} =c_{dL,A} =c_{dR,A}=1$. 
This  compares favourably to the
 CMS bound of  $\Lambda > 950$ GeV, 
for
$C_{uL,A} =C_{uR,A} =C_{dL,A} =C_{dR,A}=1$. 
Since the  analytical estimate
is reasonable,  
 most of the  dark matter signal 
 probably  comes from   $M_{inv}^2 \gg m_Z^2$, 
and the approximation  (\ref{annuler}) is consistent. 
However, the analytic bound 
is a bit to restrictive (perhaps
in part because it  includes any number of jets),
so in the remainder of the paper, the CMS
limit of 950 GeV will be used.

There is also an upper limit on the  $C$s which a
collider can exclude,   eqn (\ref{plage}),
 from requiring that
the  contact interaction approximation
be self-consistent: $C/\Lambda^2 <4\pi/M_{inv}^2$.
Since the previous  analytic estimate reproduces the
CMS bound for $M_{inv}^2 \sim$ TeV$^2$, 
the consistency condition is taken as $C< 4\pi$. 
For the axial $\chi$  current  with $\Lambda = $ TeV, 
the CMS limit  and eqn  (\ref{plage})
give   3 independent bounds on
 $\{ c_{qL,A},  c_{uR,A}, c_{dR,A}\}$:
\bea
4\pi  \lsim 
\sqrt{\frac{2}{3} |C_{qL,A}+   \frac{2}{15}
 C_{Z,A}|^2 + 
\frac{1}{3}|C_{qL,A} -   \frac{1}{6}
 C_{Z,A}|^2}
 \lsim  \sqrt{2} \nonumber\\
4\pi  \lsim 
|C_{uR,A} -   \frac{1}{15}
 C_{Z,A}|
 \lsim  \sqrt{3} ~~~~~~~~~~~~~~\nonumber\\
4\pi  \lsim 
|C_{dR,A} +   \frac{1}{30}
 C_{Z,A}|
 \lsim  \sqrt{6}~~~~~~~~~~~~~~ 
\label{lim3}
\eea
where 
 the first line is the summed contributions
of $u_L$ and $d_L$, the fractions are
approximations $gg_X^Q s_{\rm w}/2c_{\rm w}$,
 and the $d$ to $u$   pdf
ratio is taken 1/2. Similar limits
apply for the vector operator  of eqn (\ref{qDMd6}).

It can be seen already  from eqn (\ref{lim3}),
that including the interactions with the $Z$
will make little differences to the LHC limits
on the $C_{QX,A}$: for the doublet
quarks, the $Z$ contribution  cannot
cancel simultaneously against  the $u_L$ and $d_L$
contributions, and the $Z$ contribution
is irrelevant for the singlet quarks, because
also $C_{Z,A}$ must be $\lsim 4\pi$. The parameters
ruled out by the  first and second eqns of (\ref{lim3}) are
represented as the central regions in figure
\ref{fig2}.

\section{From the TeV to the MeV}
\label{sec:DD}

In direct detection, the
dark matter  scatters non-relativistically off nuclei. 
Therefore,  to translate   the EFT from the TeV
to the MeV,  the $Z$  must be removed,
the effects of QCD loops in running the operator
coefficients should be
included, and the quarks must be embedded in the  nucleons.

To remove the $Z$, the   Greens
function  for two quarks and two $\chi$s in
the effective theory with a $Z$, should be
matching to the same Greens function  in 
the theory without a $Z$. 
 Since the matching is performed at zero momentum for the
fermion legs, the contact interactions of eqn (\ref{inteff})
do not contribute, and the  coefficients of
the four-fermion operators of eqn  (\ref{qDMd6})
remain the same after the $Z$ is ``matched out''. 
The $Z$ vertices were taken $\propto p_Z^2$ to obtain this.

The light quark  currents $\overline{q} \g^\mu P_X q$ are
conserved in QCD, so  do not run. 
Also, since  $\chi$ is a  SM gauge
singlet and the only dark sector particle below the
TeV,  I suppose that  the   operators 
with vector and axial vector  $\chi$ currents
do not mix below the TeV. See {\it e.g.}
\cite{UliSDSI}  about
 loop effects mixing various operators involving
dark matter and the SM.

Finally, the quark currents can be embedded in nucleons
$N = \{p,n\}$ using  identities \cite{BBPS} such as
\bea
\langle N|  \overline{Q_i} \g^{\mu} Q_i |N \rangle 
&=& c_{V,i}^{N} \langle N| \overline{\psi_N} \g^{\mu}\psi_N  |N \rangle 
\nonumber
\eea
where $ c_{V,u}^{p}= c_{V,d}^{n} = 2$, and $c_{V,d}^{p} = c_{V,u}^{n} = 1$,
because this current counts valence quarks in the nucleon. 
The axial quark current is proportional to the nucleon
spin:
\bea
\langle N|  \overline{Q_i} \g^{\mu}\g_5 Q_i |N \rangle 
&=& 2 s^\mu  \Delta Q_i^N     = \Delta Q_i^N \langle N|  \overline{\psi_N} \g^{\mu}\g_5 \psi_N |N \rangle
\nonumber
\eea
where the proportionality  constants are measured \cite{DqN} as
$ \Delta u^p = \Delta d^n = 0.84$, $\Delta d^p = \Delta u^n = -0.43$.
In the zero-momentum-transfer limit of
non-relativistic scattering, the dark matter
can have spin-dependent interactions via
the axial current, or spin-independent  interactions
via the first component of the vector current.

The spin-independent scattering amplitude
for $\chi$ on a nucleon, is a coherent
sum of vector and scalar  interactions,
for quarks of both chiralities and all
flavours.  
The experimental limit on the cross-section per nucleon
is  $\sigma_{SI}\lsim 10^{-44}$ cm$^2$ for
$m_\chi \sim 100$ GeV \cite{Xenon}. 
For  the proton ($C_{uR}\leftrightarrow C_{dR}$ for
the neutron), with 
$ C_{qR,V} =\frac{1}{3}( C_{dR,V } + 2C_{uR,V})$,
this  gives \cite{BBPS}
\bea
\sigma_{SI} \simeq \!
\frac{1}{\pi}\! 
\left[ \frac{3m_N}{2\Lambda^2} 
 (C_{qL,V}  +  C_{qR,V} + ...)\right]^2
\!\!\! 
\lsim 3 \times 10^{-17} {\rm GeV}^{-2} 
\nonumber
\eea
where
the $+...$ contains scalar contact interactions
neglected in this note.
For $\Lambda =$ TeV,  this gives
\beq
 [C_{qL,V}  +  \frac{1}{3}( C_{dR,V } + 2C_{uR,V})
+ ...] \lsim   10^{-2} ~~~(SI).
\label{SIbd}
\eeq
The spin dependent cross-section per proton  
is \cite{BBPS}
$$
\sigma_{SD} \!\simeq   \!  m_p^2  \left[
 \frac{.42 (C_{qL,A} \!+ \!C_{uR,A}  \!- \ \!  
2C_{dR,A}  )
}{2\Lambda^2} \right] ^2 \! \! \! \lsim \frac{ 10^{-10 }}{4} {\rm GeV}^{-2}
$$
where the experimental bound is  for
$m_\chi \sim 100$ GeV.
For $\Lambda = $ TeV, this gives 
\bea
| (C_{qL,A}+C_{uR,A} -  
2C_{dR,A}  )| \lsim  20
~~~(SD).
\label{SD}
\eea
Comparing to eqn (\ref{lim3}) shows 
that the  contact interactions 
explored by SD direct detection experiments
are  mediated by
physics which is not a contact interaction
at the LHC, so are not excluded by
the limits given in eqn (\ref{lim3}).
The limit  (\ref{SD}) is represented in figure
\ref{fig2} as the vertical exclusions.

\section{Discussion}

From a bottom-up EFT point of view, it is important to include
all operators which can interfere, when  computing  experimental constaints. 
This is to allow for cancellations.
Including several operators which do not interfere
improves the bound, but is not otherwise motivated. 
In this note,   operators with 
 vector and
axial vector currents for the dark matter fermion
 $\chi$  were presented as an example,
which  illustrates two points.

\begin{figure}[ht]
\begin{center}
\epsfig{file=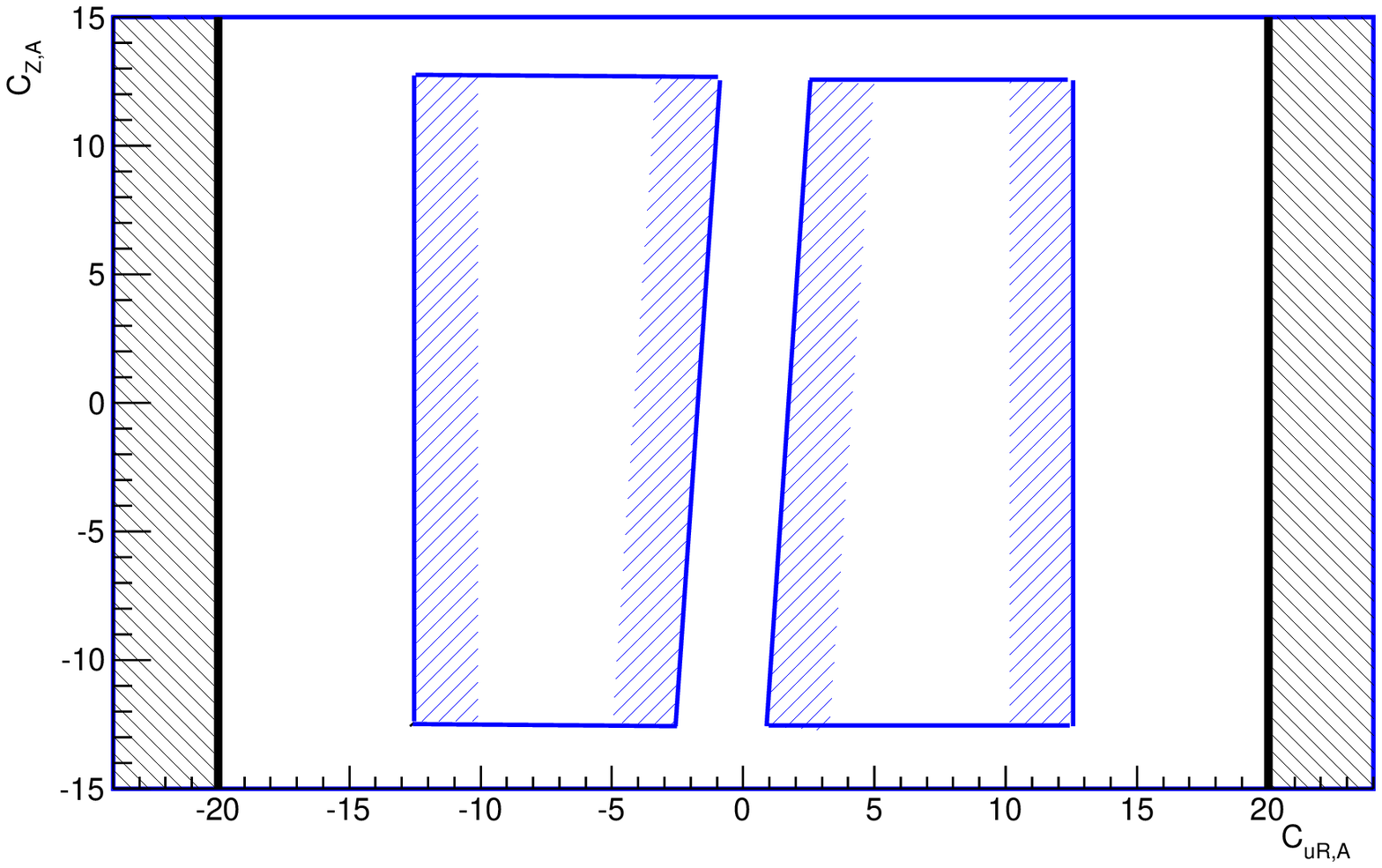, height=6.5cm,width=8.5cm}
\epsfig{file=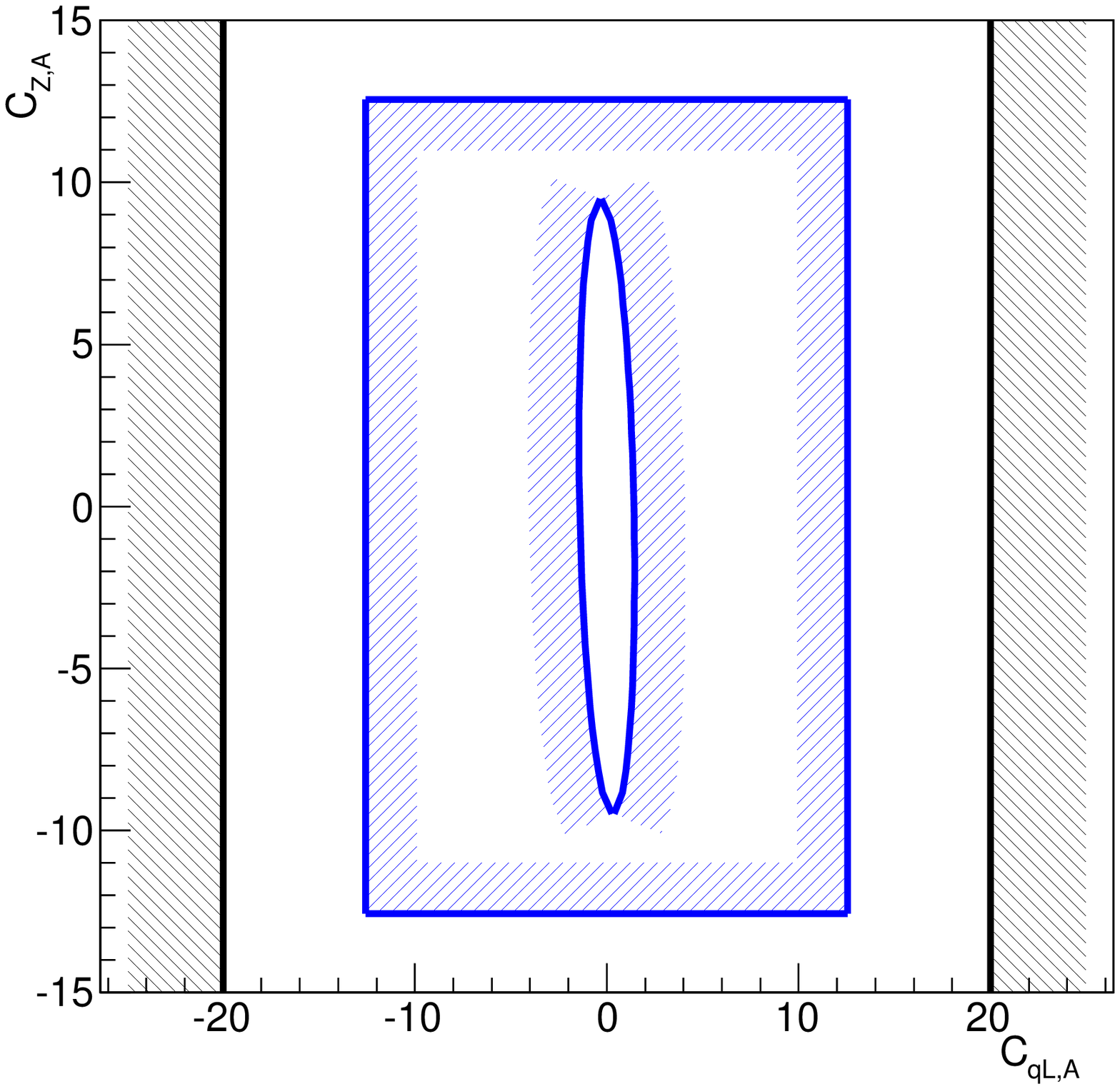, height=6.5cm,width=8.5cm}
\end{center}
\caption{ Parameter space 
excluded by
spin dependent direct detection experiments (at either side),
and the LHC (central region), for a dark matter
fermion  of mass $\sim $ 100 GeV, with
contact interactions  
with the $Z$  parametrised by
 $C_{Z,A}$  (see eqn (\ref{inteff})),
and with  $u_R$ quarks in
the left plot, and the doublet $q_L$ in the right  plot 
(see eqn  (\ref{qDMd6})).
  $\Lambda = $ TeV,  and all other  coefficients  are  zero. 
The upper
limit of the LHC exclusions is estimated from eqn (\ref{plage}). }
\label{fig2}
\end{figure}

First, the EFT at the LHC 
contains more particles   than 
 the light partons and dark matter that
are relevant in direct detection. At the LHC, the Higgs
and $Z$ should also be included.  Matching the high and low energy
EFTs, as done in this note, suggests that
the LHC constrains several
combinations  of operator
coefficients that  are different from direct detection,
as can be seen by
comparing  eqns
(\ref{lim3}) and (\ref{SD}).
However, the contribution of the $Z$ is relatively
unimportant, because its couplings to singlet
quarks are small, and it interferes with opposite
sign with $u_L$ and $d_L$.  The LHC limits
on the  dark matter couplings to quarks
and the $Z$ are represented as the
central exclusion  areas of figure \ref{fig2}:
the coupling
to quarks is more constrained  than the coupling to the $Z$,
and arbitrary  axial current  dark matter 
interactions to quarks cannot be allowed by tuning the dark matter
coupling to the $Z$.  This is because there is a
self-consistency upper bound on contact interaction
coefficients at colliders $C/\Lambda^2 < 4\pi/\hat{s}$ (see
eqn (\ref{plage})). It is important to notice
that this  upper bound also implies that
the LHC limits do not exclude the parameter
space probed by spin dependent direct detect experiments. 

Second, an interesting difference between direct  detection and
collider experiments, is that quarks of different
chirality and flavour   interfere in direct detection,
whereas the LHC can
constrain the interactions of dark matter with
each flavour and chirality of  quark individually. 
This is related to the relative
unimportance of the $Z$: it cannot cancel
separately against the contributions
of $u_L, d_L, u_R$ and $d_R$.

In summary, the rules of bottom-up Effective Field
Theory say that one should include all operators
up to some specified dimension. So to parametrise 
at dimension six 
the  axial vector interactions of dark matter with quarks, one
should include  contact interactions
of dark matter with the  quarks and with the $Z$.
Including  interactions  with the $Z$
that are $\propto p_Z^2$, as done here,
suggests that these are  not crucial. 

\subsection*{Acknowledgements}
I thank  J.P. Chou, S Malik, and S. Perries for useful comments.

\end{document}